# Thermoelectric properties of $Sr_{0.61}Ba_{0.39}Nb_2O_{6-\delta}$ ceramics annealed in different oxygen-reduction conditions

Y. Li , J. Liu[a], C.L. Wang, W.B. Su, Y.H. Zhu, J.C. Li and L.M. Mei

School of Physics, State Key Laboratory of Crystal Materials, Shandong University, Jinan 250100, P. R. China

**Abstract**: The thermoelectric properties of $Sr_{0.61}Ba_{0.39}Nb_2O_{6-\delta}$ ceramics, reduced in different conditions, were investigated in the temperature region from 323 K to 1073 K. The electrical transport behaviors of the samples are dominated by the thermal-activated polaron hopping, the Fermi glass behavior, and the Anderson localized behavior from low temperatures to high temperatures, respectively. The lattice thermal conductivity presents a plateau at high temperatures, indicating a glass-like thermal conduction behavior. Both the thermoelectric power factor and the thermal conductivity increase with the increasing degree of oxygen-reduction. Taking these two factors into account, the oxygen-reduction can still contribute to promoting the thermoelectric figure of merit. The highest *ZT* value (~0.19 at 1073 K) is obtained in the heaviest oxygen reduced sample.

[a] Corresponding author. E-mail address: liujjx@sdu.edu.cn (Jian Liu)
postal address:   School of Physics, Shandong University, Jinan, Shandong, P. R. China, 250100
telephone number: +86-531-88377035-8320; fax: +86-531-88377032

1. Introduction

Thermoelectric materials can be used to convert thermal energy and electrical energy to one another, which make sense for environmental friendly refrigeration and waste heat recycling. The thermoelectric conversion efficiency is evaluated by the thermoelectric figure of merit $ZT=S^2T/\rho\kappa$, where S is the Seebeck coefficient, T is the absolute temperature, ρ is the electrical resistivity, and κ is the thermal conductivity. Although heavy-metal-based thermoelectric materials, such as $(Bi,Sb)_2(Te,Se)_3$, SiGe, PbTe, and related semiconductors, have been well studied [1-7], there still exists some limitations, for instance, high costs, instability at high temperatures, and depending on rare or toxic elements [8-10]. Oxide thermoelectric materials can overcome these problems [8, 9], but the thermoelectric properties of oxide are still far from practical use. Therefore, it is significant to find new oxide thermoelectric materials with a high *ZT* value.

Recently, tungsten bronze-structured ferroelectric $Sr_{1-x}Ba_xNb_2O_{6-\delta}$ (SBN) was found to be a promising n-type thermoelectric oxide. It was observed that $Sr_{0.61}Ba_{0.39}Nb_2O_{6-\delta}$ (SBN61) single crystals have high thermoelectric power factors (~2000 μW/mK$^2$ at 516 K) parallel to the c-axis [10]. In the SBN structure, the Nb-O octahedrons are interconnected orderly along the c-axis, and show a complex arrangement in the ab plane. There exist many vacancies among the Nb-O octahedrons, occupied by $Sr^{2+}$ or $Ba^{2+}$ ions. This complex system has formerly been concerned with its low thermal conductivity [10-12], which may contribute to promoting the thermoelectric performance. In the oxygen reduced SBN61 ceramics, the concentration of oxygen vacancies increases through the oxygen-reduction treatment, and so do the electron concentration and the ratio of $Nb^{4+}/Nb^{5+}$. The negative charge carriers contribute to the electrical conductivity, and the oxygen vacancies and $Nb^{4+}$ ions may influence the thermal conduction [10]. As a new thermoelectric oxide, there are many questions that still need to be investigated. For example, the thermal conductivity and exact *ZT* value of reduced SBN61 ceramics has not been reported yet, especially at high temperatures. In this paper, we measured the thermoelectric properties of the oxygen reduced $Sr_{0.61}Ba_{0.39}Nb_2O_{6-\delta}$ ceramics from 323 K to 1073 K. The electrical transport mechanism and the thermal conduction behavior were analyzed detailedly. The influence on thermal properties of oxygen-reduction was discussed, and it concluded that the heavily oxygen-reduction effectively promotes the thermoelectric figure of merit for SBN61 ceramics.

2. **Experimental**

$Sr_{0.61}Ba_{0.39}Nb_2O_{6-\delta}$ ceramics were prepared by solid-state reaction techniques using $SrCO_3$ (99.8%), $BaCO_3$ (99.8%), and $Nb_2O_5$ (99%) powders. Appropriate amounts of the starting materials were ground and pressed into pellet discs. These discs were calcined at 1200 °C for 6 hours in an air atmosphere. After intermediate grinding and pressing, the discs were sintered at 1350°C in air atmosphere. And then, the discs were annealed at 1100 °C or 1250 °C for 10 hours in pure argon or forming gas (5 mol% hydrogen in argon) with a flow rate of 0.2 *l*/min. The phase structures were investigated by powder x-ray diffraction (XRD) with a Rigaku D/MAX-2550P diffractometer using Cu *Kα* radiation (λ=0.154056 nm). Scanning electron microscope (SEM, Hitachi SU-70) equipped with energy dispersive spectrometer (EDS) was applied for the surface microstructure and composition evaluation. The thermal conductivity was calculated from the thermal diffusivity, the specific heat capacity, and the sample density as measured on a laser flash apparatus (Netzsch LFA 427), a thermal analyser (Netzsch STA 449C), and by the Archimedes' method, respectively. The sintered discs were cut into rectangular columns (16×2×2 $mm^3$) for measurements of the Seebeck coefficient and the electrical conductivity using a Linseis LSR-3/1100 instrument in helium atmosphere by a modified dynamic method.

## 3. Results and Discussion

### 3.1. X-ray Diffraction and SEM Analysis

Fig. 1 shows X-ray diffraction data of the SBN61 ceramics annealed in various oxygen-reduction conditions. Both the reductive forming gas and the high annealing temperature contribute to deepen the oxygen-reduction degree, therefore, the oxygen-reduction degree increases in the order of SBN61Ar (annealed at 1250°C in argon), SBN61H1100 (annealed at 1100°C in gas mixtures of $H_2/Ar$), and SBN61H1250 (annealed at 1250°C in gas mixtures of $H_2/Ar$). The XRD patterns show that the profiles of every sample include all the diffraction peaks of tungsten bronze-structured $Sr_{0.61}Ba_{0.39}Nb_2O_6$, and a $NbO_2$ second phase was found in the heavily oxygen reduced sample SBN61H1250. This might be because that the content of $Nb^{4+}$ ions is increased by oxygen-reduction reaction, resulting in the generation of the $Nb^{4+}O_2$ second phase [11].

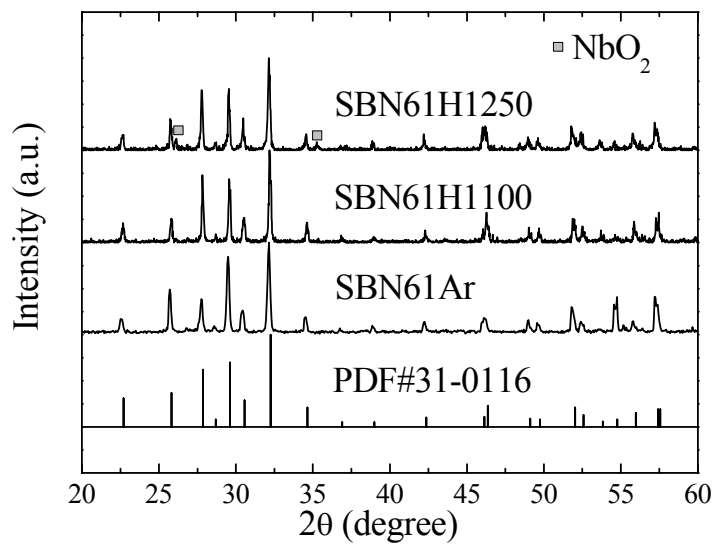

Fig. 1 XRD patterns for SBN61 ceramics. The samples were annealed at 1250°C in argon (SBN61Ar), 1100°C in gas mixtures of $H_2/Ar$ (SBN61H1100), and 1250°C in gas mixtures of $H_2/Ar$ (SBN61H1250). The $NbO_2$ second phase is marked by solid squares in the figure.

Table I. Lattice parameters (*a, b, c*), theoretical densities ($\rho_{th}$), measured densities ($\rho_m$) and relative densities ($\rho_r$) of the samples.

|  | a=b(Å) | c(Å) | $\rho_{th}$ (g/cm$^3$) | $\rho_m$ (g/cm$^3$) | $\rho_r$ |
|---|---|---|---|---|---|
| SBN61Ar | 12.441 | 3.931 | 5.31 | 5.14 | 96.8% |
| SBN61H1100 | 12.434 | 3.921 | 5.33 | 5.18 | 97.2% |
| SBN61H1250 | 12.466 | 3.928 | 5.29 | 5.14 | 97.2% |

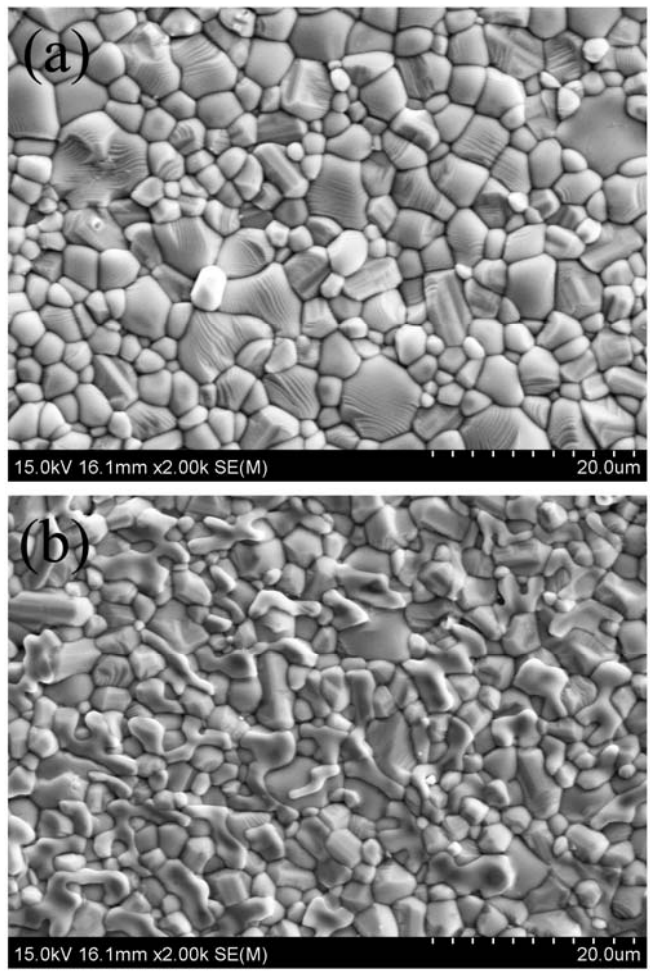

Fig. 2 SEM micrographs of sample SBN61Ar (a) and sample SBN61H1250 (b).

According to the results of XRD, the lattice parameters and theoretical densities of the samples were calculated as shown in Table I. The lattice parameters do not vary monotonically with the increasing degree of oxygen-reduction. The non-monotonic variance of the lattice parameters may be affected by many factors, such as the oxygen vacancies, the large radius of $Nb^{4+}$, and the bond angles in the $NbO_6$ octahedrons. The relative densities were calculated, also shown in Table I, and all the samples have high relative densities. Figure 2 shows the SEM micrographs of samples SBN61Ar and SBN61H1250. As shown in this figure, the SBN61 ceramics have compact microstructures in no matter heavily or slightly oxygen reduction degree. As shown in Fig. 2(b), there are two kinds of grains: coarse grains and slender grains in the heavily reduced sample. According to the results of EDS, the coarse grains are the SBN61 crystal and the slender grains are the $NbO_2$ second phase. The slender grains do not exist in the sample SBN61Ar as shown in Fig. 2(a), which is consistent with the XRD results.

3.2. Seebeck coefficient and Electrical Resistivity

Figure 3 shows the electrical resistivity $\rho$ and the Seebeck coefficient $S$ for samples SBN61Ar, SBN61H1100, and SBN61H1250, which intensively measured below 623 K. The Seebeck coefficient below temperature 508 K shown in Fig. 3(a) reflects a similar behavior with that reported by Lee *et al*. [11]. Moreover, we measured the Seebeck coefficient and the electrical resistivity up to 1073 K and found a different transport behavior in the high temperature region (above 508 K). As shown in Fig. 3(a) and (b), both the electrical resistivity and the magnitudes of Seebeck coefficient decrease with the deepening oxygen-reduction degree. As a result, the thermoelectric power factor ($PF=S^2/\rho$) is enhanced from 182 $\mu W/K^2 m$ (in SBN61Ar at 1073 K) to 375 $\mu W/K^2 m$ (in SBN61H1250 at 1073 K).

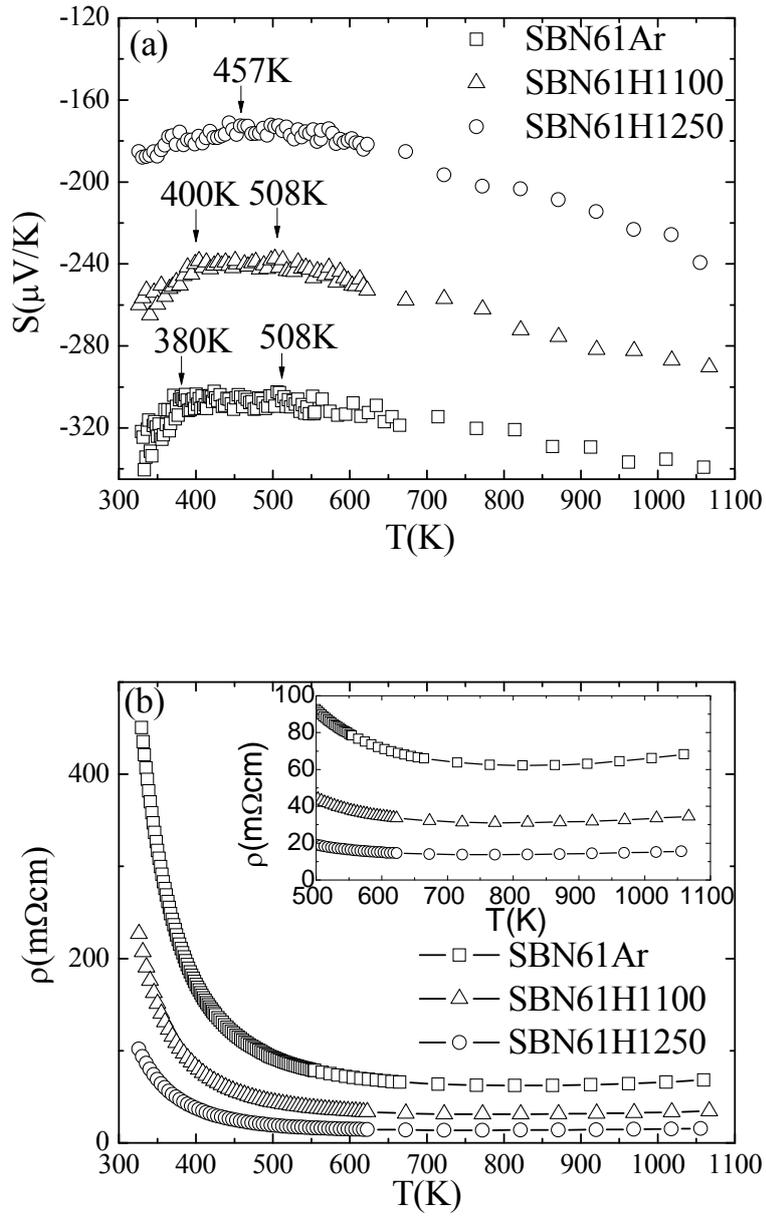

Fig. 3 *S* versus T (a) and *ρ* versus T (b) plots for samples SBN61Ar, SBN61H1100, and SBN61H1250. The insert figure in Fig. 3(b) shows the *ρ* versus T plots in the temperature range from 500 K to 1100 K.

In the following, the electrical transport properties were discussed in detail. The Seebeck coefficient directly reflects the behavior of the band structure and is seldom influenced by the scattering processes at high temperatures (above room temperature), therefore, the discussion is mainly based on the behavior of Seebeck coefficient. As shown in Fig. 3, the behavior of Seebeck coefficient of the SBN61Ar and SBN61H1100 samples can be divided into three parts: part I is in the low temperature region (below 380 K for SBN61Ar, and below 400 K for SBN61H1100); part II is in the middle temperature region (380 K~508 K for SBN61Ar, and 400 K~508 K for SBN61H1100); part III is in the high temperature region (above 508 K for SBN61H1100 and SBN61Ar). And that of the SBN61H1250 sample can be divided into two parts by the temperature 457 K.

Firstly, the slightly reduced samples SBN61Ar and SBN61H1100 are discussed. For these two samples, they both have three parts. In part I, the magnitudes of Seebeck coefficient decrease with the increasing temperature, and the $\ln\rho$ for the two samples show linear dependence on 1/T as shown in Fig. 4(a). It indicates a thermal activation behavior, and $\rho$ and $S$ can be described by [13]:

$$\rho = \rho_0 \exp(E_\rho / k_B T) \quad (1)$$

$$S = (k_B/e)\exp(E_s/k_B T + A) \quad (2)$$

where $\rho_0$ and A are constants, $k_B$ is Boltzmann constant, $E_\rho$ is the overall activation energy for conduction, and $E_s$ is the activation energy associated with carrier generation. From the Eq. 1, Eq. 2 and the data in Fig. 4(a), the $E_\rho$ and the $E_s$ for sample SBN61Ar and sample SBN61H1100 can be calculated: $E_\rho$ is about 0.29eV, and $E_s$ is about 0.01eV. Apparently, $E_\rho$ is larger than $E_s$, so the transport behavior in part I is dominated by the thermal activated polaron hopping [13, 14].

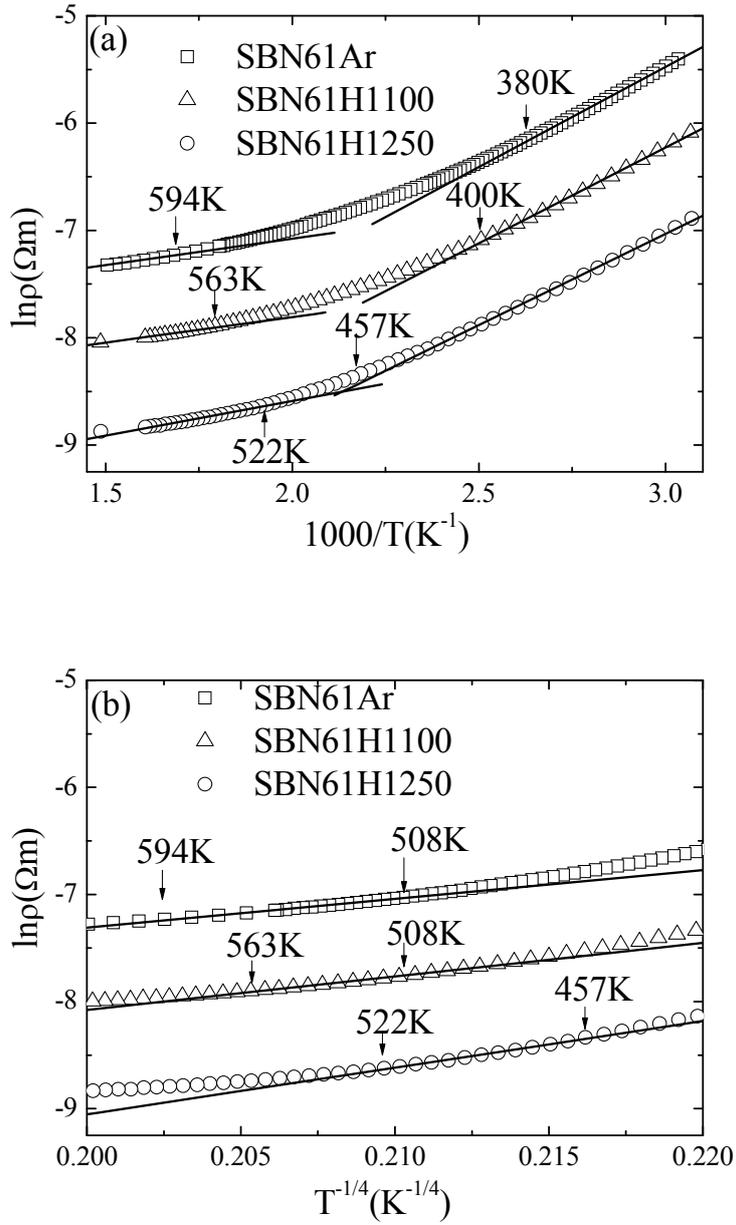

Fig. 4 ln$\rho$ versus 1000/T (a) and ln$\rho$ versus $T^{-1/4}$ (b) plots for samples SBN61Ar, SBN61H1100, and SBN61H1250.

In part II, it shows a temperature-independent S for the two samples SBN61Ar and SBN61H1100 in Fig. 3(a). There is an obvious change in Seebeck coefficient from part I to part II, which probably correlated with the 4*mm* to 4/*mmm* ferroelectric phase transition [10, 11]. In this part, the temperature-independent Seebeck coefficient can be described by: $S = -(k_B/e)\ln[(2-c)/c]$, where c is the ratio of the number of electrons to the number of sites [15]. It indicates that the Seebeck coefficient is determined solely by the constraints as to the occupancy of states and the density of electrons in the system, and the temperature independence of Seebeck coefficient is governed entirely by the entropy change per added carrier [15, 16]. Considering the complex crystal structure and large disorder in SBN61, this behavior is probably due to the Fermi glass behavior, in which the disorder is great enough to ensure that electrons are localized in the Anderson sense [13]. However, the behavior of electrical resistivity in this temperature range is neither completely consistent with the Fermi glass model, nor with other classical insulating models (such as the thermal activated model or variable range hopping model). It may be due to the reason that the electrical resistivity becomes difficult to get good fitting effects when the temperature increases into the region of the relaxor ferroelectric phase transition.

As shown in Fig. 3(a), it also shows a Seebeck coefficient anomaly from part II to part III for the two samples SBN61Ar and SBN61H1100, however, what causes this anomaly still remains to be investigated. In part III, the electrical resistivity of these two samples indicates a transition from semiconductor behavior to metal-like behavior at about 770 K, as shown in Fig. 3(b). However, the magnitudes of *S* increase steadily with the increasing temperature, and the slope of *S* versus T plot does not change obviously. Taking these factors into account, the transport properties are anticipated to be dominated by the Anderson localized behavior in a conduction band. Therefore, the resistivity should follow the variable range hopping (VRH) model at first, and the relation of ρ and T is given by Mott's law [17]:

$$\rho = \rho_1 \exp[C(T_0/T)^{1/4}] \qquad (3)$$

where $\rho_1$ and C are constants, and $T_0$ is Mott characteristic temperature. As shown in Fig. 4(b), the ln$\rho$

linearly depends on $T^{-1/4}$ plot at the first temperature region in part III, following the VRH model. Then, when the temperature is above $T_0$, the electrons begin to enter into the extended states, and the hopping turns to the fixed range hopping. Thus, the resistivity is decided by the thermal activation behavior, and the relation of $\rho$ and T follows the equation [17]:

$$\rho = \rho_2 \exp(\Delta\varepsilon/k_B T) \qquad (4)$$

where $\rho_2$ is a constant, and $\Delta\varepsilon=\varepsilon_c-\varepsilon_F$ is thermal activation energy (where $\varepsilon_c$ is the mobility edge, and $\varepsilon_F$ is the Fermi level). The ln$\rho$ linearly depending on 1/T plot above $T_0$ (~594 K for SBN61Ar, 563 K for SBN61H1100) is also shown in Fig. 4(a). Finally, because that plenty of electrons enter into the extended states, no more electrons enter into the extended states from the localized states with the increasing temperature, and the scattering plays an important role in the electrical resistivity. In this case, the resistivity should follow the metal-like model: $\rho \sim T^\alpha$, where α is an exponent of T [17]. As shown in Fig. 3(b), $\rho$ begins to increase with the increasing temperature above 770 K. In conclusion, the electrical transport behavior in part III for SBN61Ar and SBN61H1100 is consistent with the Anderson localized theory.

Then, we turned to discuss the heavily oxygen reduced sample SBN61H1250. As shown in Fig. 3(a), the behavior of *S* of SBN61H1250 could be divided into two parts. The plateau *S* in the middle temperature region was not observed in this sample. Using the same methods discussed above, it could be concluded that the electrical transport mechanism of SBN61H1250 also follows thermal-activated polaron hopping (below 457 K) and the Anderson localized behavior (above 457 K), respectively.

From the discussion above and the data in Fig. 3, it shows that different oxygen-reduction degree has some influences on the electrical transport behavior. As shown in Fig. 3(a), the middle temperature region with the plateau Seebeck coefficient of sample SBN61Ar is longer than that of the sample

SBN61H1100, and the plateau *S* does not appear in the sample SBN61H1250. There are many possible reasons for this transition, such as the effects of the concentration of oxygen vacancy and $Nb^{4+}$ ions, or the influence of $NbO_2$ second phase, which calls for more detailed investigation.

3.3. Thermal Conductivity

Figure 5 shows the total thermal conductivity $\kappa$ (a), and the lattice thermal conductivity $\kappa_l$ (b) as a function of temperature. $\kappa_l$ was calculated by subtracting the electronic thermal conductivity $\kappa_e$ from $\kappa$. And $\kappa_e$ is calculated by the Wiedemann–Franz relation, $\kappa_e = L\sigma T$ (where $L=2.23\times 10^{-8}$ $V^2/K^2$ is Lorenz number, $\sigma$ is electrical conductivity, and T is absolute temperature) [17]. The $\kappa_l$ of SBN61H1100 and SBN61H1250 is higher than the $\kappa_l$ of SBN61Ar, but the $\kappa_l$ of SBN61H1100 and SBN61H1250 is very close to each other. It indicates that the oxygen-reduction can increase the lattice thermal conductivity of the SBN61 ceramics. However, when the oxygen-reduction achieves some degree, the deepening of the oxygen-reduction degree has little effect on the increase of the lattice thermal conductivity. As shown in Fig. 5(b), the lattice thermal conductivity of these three samples exhibits the same behavior. $\kappa_l$ increases with the increasing temperature at first, and then, the high-temperature plateau $\kappa_l$ (~1.77 W/Km, 1.97 W/Km, 1.98 W/Km for SBN61Ar, SBN61H1100, and SBN61H1250, respectively) is observed when the temperature is above 530 K. This thermal conduction behavior indicates a glass-like thermal conductivity [18, 19]. The increase of the lattice thermal conductivity with the increasing oxygen-reduction degree leads to the conclusion that the defect scattering has little influence on this glass-like thermal conduction behavior [18]. However, what mainly influences the thermal conductivity still needs to be investigated.

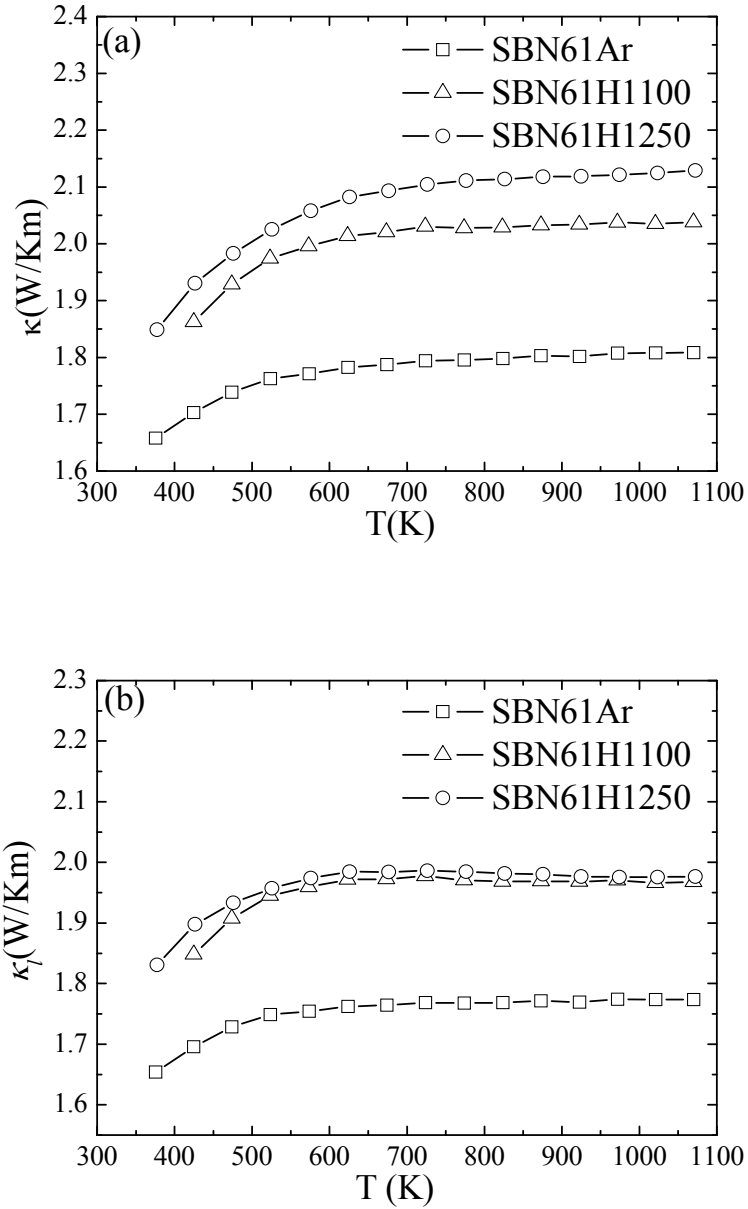

Fig. 5 Total thermal conductivity (a), lattice thermal conductivity (b) as a function of temperature for samples SBN61Ar, SBN61H1100, and SBN61H1250.

3.4. Thermoelectric Figure of Merit

The thermoelectric figure of merit (*ZT*) of SBN61Ar, SBN61H1100, and SBN61H1250 as a function of temperature is shown in Fig. 6. The *ZT* values for all the samples increase with the increasing temperature in the whole temperature range. Although, as discussed above, the oxygen-reduction raises the thermal conductivity, the increase of the *PF* value plays a more important role in the thermoelectric figure of merit. Therefore, a higher *ZT* value is obtained in the SBN61 ceramics reduced in a deeper degree, and the sample SBN61H1250 exhibits the highest *ZT* value (~0.19 at 1073 K). It can be concluded that the oxygen-reduction is helpful to promote the thermoelectric properties of SBN61 ceramics.

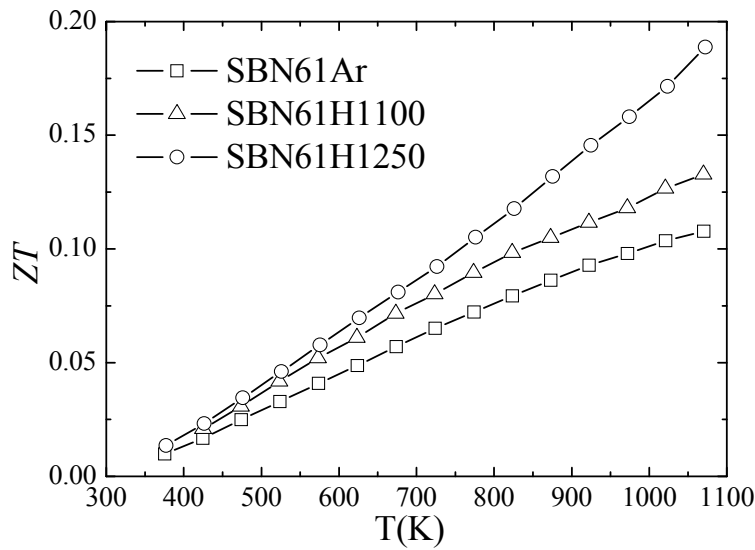

Fig. 6 Thermoelectric figure of merit (*ZT*) as a function of temperature for samples SBN61Ar, SBN61H1100, SBN61H1250.

## 4. Conclusion

The thermoelectric properties of $Sr_{0.61}Ba_{0.39}Nb_2O_{6-\delta}$ ceramics, reduced in different conditions, were investigated in the temperature range from 323 K to 1073 K. In the slightly oxygen reduced $Sr_{0.61}Ba_{0.39}Nb_2O_{6-\delta}$ ceramics, the electrical transport mechanism follows a thermal activated polaron hopping behavior when the temperature is below 380 K. Then the Seebeck coefficient becomes invariant with the temperature in the temperature range between 380 K and 508 K, which probably due to a Fermi glass behavior. When the temperature is higher than 508 K, the electrical transport behaviors are dominated by the Anderson localized transport behavior in a conduction band. In the deeper oxygen reduced $Sr_{0.61}Ba_{0.39}Nb_2O_{6-\delta}$ samples, the plateau Seebeck coefficient in the middle temperature range shortens with the increasing oxygen-reduction degree, and disappears in the heaviest reduced sample. The absolute Seebeck coefficient and the electrical resistivity decrease with the deepening oxygen-reduction degree, however, the electrical resistivity plays a major role in the thermoelectric power factor (PF) of the reduced SBN61 ceramics. Therefore, the *PF* values increase with the deepening oxygen-reduction degree. The lattice thermal conductivity of all the reduced SBN61 samples shows a glass-like behavior. Although the oxygen-reduction increases the thermal conductivity of SBN61 ceramics, it still can dramatically enhance the *ZT* value via promoting the thermoelectric power factor. And the heaviest oxygen reduced sample SBN61H1250 exhibits the highest *ZT* value (~0.19 at 1073 K).

**Acknowledgements**

The authors acknowledge the financial support of the National Basic Research Program of China (2013CB632506) and Natural Science Foundation of China (51202132 and 11374186).

**References**

[1] G.J. Snyder, M. Christensen, E. Nishibori, T. Caillat, B.B. Iversenet, Disordered zinc in $Zn_4Sb_3$ with phonon-glass and electron-crystal thermoelectric properties, Nat. Mater. 3 (2004) 458-463.

[2] T.E. Svechnikova, L.E. Shelimova, P.P. Konstantinov, M.A. Kretova, E.S. Avilov, V.S. Zemskov, C. Stiewe, A. Zuber, E. Muller, Thermoelectric properties of $(Bi_2Te_3)_{1-x-y}(Sb_2Te_3)_x(Sb_2Se_3)_y$ single crystals, Inorg. Mater. 10 (2005) 1043-1049.

[3] B.C. Sales, D. Mandrus, R.K. Williams, Filled Skutterudite Antimonides: A New Class of Thermoelectric Materials, Science 272 (1996) 1325-1328.

[4] G.S. Nolas, J.L. Cohen, G.A. Slack, S.B. Schujman, Semiconducting Ge clathrates: Promising candidates for thermoelectric applications, Appl. Phys. Lett. 73 (1998) 178.

[5] R. Venkatasubramanian, E. Siivola, T. Colpitts, B. O'Quinn, Thin-film thermoelectric devices with high room-temperature figures of merit, Nature 413 (2001) 597-602.

[6] T.M. Tritt, Thermoelectric Materials: Principles, Structure, Properties, and Applications, in: Encyclopedia of Materials: Science and Technology (2002), pp. 1-11.

[7] K.F. Hsu, S. Loo, F. Guo, W. Chen, J.S. Dyck, C. Uher, T. Hogan, E.K. Polychroniadis, M.G. Kanatzidis, Cubic $AgPb_mSbTe_{2+m}$: Bulk Thermoelectric Materials with High Figure of Merit, Science 303 (2004) 818-821.

[8] H. Ohta, Thermoelectrics based on strontium titanate, Mater. Today 10 (2007) 44-49.

[9] I. Terasaki, High-temperature oxide thermoelectrics, J. Appl. Phys. 110 (2011) 053705.

[10] S. Lee, R.H.T. Wilke, S. Trolier-McKinstry, S.J. Zhang; C.A. Randall, $Sr_xBa_{1-x}Nb_2O_{6-\delta}$ Ferroelectric-thermoelectrics: Crystal anisotropy, conduction mechanism, and power factor, Appl. Phys. Lett. 96 (2010) 031910.

[11] S. Lee, S. Dursun, C. Duran, C.A. Randall, Thermoelectric power factor enhancement of textured ferroelectric Sr$_x$Ba$_{1-x}$Nb$_2$O$_{6-\delta}$ ceramics, J. Mater. Res. 26 (2011) 26-30.

[12] C.L. Choy, W.P. Leung, T.G. Xi, Y. Fei, C.F. Shao, Specific heat and thermal diffusivity of strontium barium niobate Sr$_{1-x}$Ba$_x$Nb$_2$O$_6$ single crystals, J. Appl. Phys. 71 (1992) 70

[13] N.F. Mott, E.A. Davis, Electronic Processes in Non-Crystalline Materials, second ed., Oxford University Press, New York, 1971.

[14] M. Jaime, M.B. Salamon, K. Pettit, M. Rubinstein, R.E. Treece, J.S. Horwitz, D.B. Chrisey, Magnetothermopower in La$_{0.67}$Ca$_{0.33}$MnO$_3$ thin films, Appl. Phys. Lett. 68 (1996) 1576.

[15] R.R. Heikes, R.W. Ure, Thermoelectricity: Science and Engineering, Interscience, New York, 1961, p. 77.

[16] P.M. Chaikin, G. Beni, Thermopower in the correlated hopping regime, Phys. Rev. B 13 (1976) 647-651.

[17] C. Kittel, Introduction to Solid State Physics, seventh ed., John Wiley, New York, 1996.

[18] E. Fischer, W. Hässler, E. Hegenbarth, Glass-Like Behaviour in the Thermal Conductivity of Sr$_{1-x}$Ba$_x$(Nb$_2$O$_6$) Single Crystal, Phys. Stat. Sol. (a) 72 (1982) 169-171.

[19] C. Kittel, Interpretation of the Thermal Conductivity of Glasses, Phys. Rev. 75 (1949) 972-974.